\documentclass{emulateapj}

\slugcomment{To appear in PASP, February 2006 issue}

\shorttitle{The Automatic Real-time GRB Pipeline of the 2-m Liverpool Telescope}
\shortauthors{Guidorzi et al.}

\begin{document}


\title{The Automatic Real-time GRB Pipeline of the 2-m Liverpool~Telescope}

\author{C. Guidorzi\altaffilmark{1},
A. Monfardini\altaffilmark{2}, A. Gomboc\altaffilmark{1,3}, C. J. Mottram,
C. G. Mundell\altaffilmark{4}, I. A. Steele, D. Carter, M.~F.~Bode\altaffilmark{5},
R. J. Smith, S. N. Fraser, M. J. Burgdorf, A.M. Newsam}
\affil{Astrophysics Research Institute, Liverpool John Moores University,
Twelve Quays House, Birkenhead, CH41 1LD, UK}

\email{(crg, am, ag, cjm, cgm, ias, dxc, mfb, rjs, snf, mjb, amn) @astro.livjm.ac.uk}


\altaffiltext{1}{Marie Curie Fellow.}
\altaffiltext{2}{present address: ITC--IRST and INFN, Trento, via Sommarive, 18 38050 Povo (TN), Italy.}
\altaffiltext{3}{present address: FMF, University in Ljubljana, Jadranska 19, 1000 Ljubljana, Slovenia.}
\altaffiltext{4}{Royal Society University Research Fellow.}
\altaffiltext{5}{PPARC Senior Fellow.}


\begin{abstract}
The 2-m Liverpool Telescope (LT), owned by Liverpool John Moores University,
is located in La Palma (Canary Islands) and operates in fully robotic mode.
In 2005, the LT began conducting an automatic GRB follow-up program.
On receiving an automatic GRB alert from a Gamma-Ray Observatory ({\em Swift},
{\em INTEGRAL}, {\em HETE-II}, {\em IPN}) the LT initiates a special override
mode that conducts follow-up observations within 2--3~min of the GRB onset.
This follow-up procedure begins with an initial sequence of short (10-s)
exposures acquired through an $r'$ band filter. These images are
reduced, analyzed and interpreted automatically using pipeline
software developed by our team called ``{\em LT-TRAP}'' ({\em Liverpool Telescope
Transient Rapid Analysis Pipeline}); the automatic
detection and successful identification of an unknown and potentially
fading optical transient triggers a subsequent multi-color imaging
sequence. In the case of a candidate brighter than $r'=15$, either a
polarimetric (from 2006) or a spectroscopic observation
(from 2007) will be triggered on the LT. If no candidate is identified,
the telescope continues to obtain $z'$, $r'$ and $i'$ band imaging with
increasingly longer exposure times. Here we present a detailed
description of the {\em LT-TRAP} and briefly discuss
the illustrative case of the afterglow of GRB~050502a, whose automatic
identification by the LT just 3 min after the GRB, led to the
acquisition of the first early-time ($<$1~hr) multi-color light curve of
a GRB afterglow.
\end{abstract}





\section{INTRODUCTION}
Since the discovery of the first optical afterglow \citep{Paradijs97}
associated with a Gamma-Ray Burst (GRB), several robotic telescopes have
been developed to perform rapid follow-up observations of GRB afterglows.
Fast robotic follow-ups are now possible because of the improved localization
of GRBs provided by current gamma-ray satellites
({\em HETE-II}, {\em INTEGRAL}, {\em Swift}, {\em InterPlanetary Network});
real-time GRB alerts are received from these satellites through the
GRB Coordinates Network (GCN) \citep{Barthelmy05} within a few seconds
of the GRB onset and provide positions with typical uncertainties
of a few arcmin.
Observations within minutes of the GRB can be crucial to determine
the properties of the close circumburst environment, potentially
providing clues to the nature of the progenitor.

\begin{figure}
\plotone{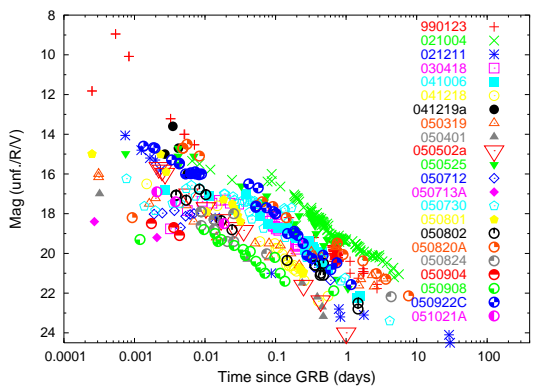}
\caption{Early light curves (unfiltered, $R$ and $V$) for a set of GRBs with detections
within minutes of the gamma--ray event. Big empty upside down triangles show the case of
GRB~050502a (filter $r'$) robotically detected and followed-up by LT
(from \citet{Guidorzi05a}).
All the measurements shown have been taken from GCN circulars
(Nos. 248, 257, 307, 715, 1572, 1589, 1596, 1732, 1738, 1739, 1748, 1752, 1759, 1821, 2773, 2776, 2784, 2798, 2799, 2821, 2875, 2897, 3118, 3165, 3480, 3485, 3487, 3489, 3492, 3575, 3584, 3588, 3596, 3601, 3604, 3646, 3704, 3705, 3711, 3712, 3716, 3717, 3718, 3720, 3723, 3726, 3728, 3733, 3736, 3739, 3741, 3744, 3745, 3746, 3756, 3765, 3775, 3778, 3829, 3834, 3836, 3838, 3853, 3863, 3864, 3865, 3868, 3869, 3870, 3880, 3883, 3896, 3907, 3917, 3929, 3943, 3944, 3945, 3947, 3950, 3953, 3960, 4011, 4012, 4015, 4016, 4018, 4023, 4026, 4027, 4032, 4041, 4044, 4046, 4048, 4049, 4095, 4120, 4121),
except for GRB~030418 \citep{Rykoff04}, GRB~041219a \citep{Vestrand05}
and GRB~050502a \citep{Guidorzi05a}.
\label{fig:LC}}
\end{figure}

To date, only a few GRBs have measurements of their optical
afterglow starting as early as a few minutes after the GRB:
990123, 021004, 021211, 030418, 041006, 041218, 041219a, 050319, 050401,
050502a, 050525, 050712, 050713A, 050730, 050801, 050802, 050820A,
050824, 050904, 050908, 050922C, 051021A (October 2005).
Figure~\ref{fig:LC} shows the early afterglow light curves for
all of them within the first day after the burst. For each GRB we
collected unfiltered, $R$, $R_c$ and $V$ observations.

\section{THE LIVERPOOL TELESCOPE}
\label{sec:LT}

The Liverpool Telescope (LT), owned and operated by Liverpool John
Moores University (LJMU), has a 2-m diameter primary mirror, altitude-azimuth
design, final focal ratio f/10. It was designed and built by Telescope Technologies
Limited, with the robotic control system and instrumentation being provided by the
LJMU Astrophysics Research Institute \citep{Steele04}.
It is situated at the Observatorio del Roque de los Muchachos, on the Canary Island La Palma.
Due to the fully opening enclosure and to the fast slew rate of 2$^{\circ}$/s,
the telescope can start observations within 1--3~min of the
receipt of the GCN alert notice. LT has five instrument ports: four folded and one
straight-through, selected by a deployable rotating mirror in the AG Box within 30~s.
The telescope is equipped with the {\em RATCam} Optical CCD Camera,
and with the {\em SupIRCam} 1--2.5~$\mu$m Camera.
A {\em Prototype Spectrograph} 
and an optical polarimeter ({\em RINGO}) based on a design by \cite{Clarke02}
are expected to be operating later in 2006.
A higher resolution, higher sensitivity and generally more capable spectrograph ({\em FRODOSpec})
is being developed for deployment in 2007.
Table~\ref{tab:instr} reports the main characteristics of each instrument.
In addition to the GRB follow-up program, which has high priority on the LT,
the robotic control \citep{Fraser02} and automated scheduler \citep{Steele97,Fraser04}
allow the LT to conduct an optimized science program of observations of time variable
sources over a wide range of timescales from Target of Opportunity (e.g. novae, supernovae,
anomalies in gravitational lenses) to longer-term monitoring (e.g. AGN, variable stars),
when no GRB observations are required.
\begin{deluxetable*}{ll}
\tabletypesize{\scriptsize}
\tablecaption{Liverpool Telescope Instrumentation\label{tab:instr}}
\tablewidth{0pt}
\tablehead{
\colhead{Instrument} & \colhead{Description}\\
}
\startdata
{\it RATCam} Optical CCD Camera - & 2048$\times$2048 pixels, 0.135"/pixel, FOV 4.6'$\times$4.6', \\
  & 8 filter selections (u', g', r', i', z', B, V, H$\alpha$, ND2.0) \\
 & - from LT first light, July 2003 \\
\hline
 {\it SupIRCam} 1 - 2.5 micron Camera - & 256$\times$256 pixels, 0.4"/pixel, FOV 1.7'$\times$1.7', \\
  (with Imperial College) & Z, J, H, K' filters - from late 2005 \\
\hline
 {\it Prototype Spectrograph} - & 49, 1.7" fibres, 512$\times$512 pixels, R=1000; \\
  (with University of Manchester) & 3500 $<$ $\lambda$ $<$ 7000 \AA - from 2006 \\
\hline
 {\it RINGO} optical polarimeter & Ring Polarimeter based on the design of \citep{Clarke02} \\
& - from 2006\\
\hline
 {\it FRODOSpec} Integral field  & R=4000, 8000; \\
  double beam spectrograph - & 4000 $<$ $\lambda$ $<$ 9500 \AA - from 2007 \\
  (with University of Southampton) & \\
\enddata
\end{deluxetable*}

\newpage
\section{REAL-TIME GRB FOLLOW-UP STRATEGY - OVERVIEW} 
When a GRB fulfills all the requirements to be observed with the LT, the GRB
over-ride mode is triggered. The GRB program takes control of the
telescope and triggers the so-called ``Detection Mode''
(Sec.~\ref{sec:DM}), during which a small number of Sloan $r'$-band
short-exposure images are taken\footnote{From end 2006, this detection mode
will be preceded by a short 30-s exposure taken with the RINGO
polarimeter.}. These images are then transferred to
the LT proxy machine, a network-attached on-site computer.  The
automatic LT GRB pipeline, called {\em Liverpool Telescope Transient
Rapid Analysis Pipeline} (hereafter {\em LT-TRAP}), is then invoked to
process, analyze and interpret these images and returns a value to
indicate whether any variable optical transient (OT) candidate has
been identified.  The result of this analysis drives the automatic
selection of the subsequent observing mode as follows: 
\begin{itemize} 
\item identification of a fading OT with $r'>15$\ $\rightarrow$ Multi-Color Imaging Mode (Sec.~\ref{sec:MCIM}); 
\item identification of a fading OT with $r'\le15$ $\rightarrow$ Polarimetry
Mode (2006)/Spectroscopy Mode (2007) (Sec.~\ref{sec:SM}); 
\item no identification of an OT $\rightarrow$ RIZ Mode (Sec.~\ref{sec:RIZ}).
\end{itemize} 
Figure~\ref{fig:strategy} shows a diagram of the
overall observing strategy when the GRB over-ride mode is triggered.

\section{LT GRB OVER-RIDE MODE}
The GRB Target of Opportunity Control Agent (GRB TOCA) is a shell script that
is invoked in response to a GRB alert being received at LT via a socket connection.
This script uses the TOCA of the LT Robotic Control System (RCS) to take control
of the telescope, provided that the GRB position is observable
and other requirements are fulfilled, e.g.:
\begin{itemize}
\item the positional error circle is sufficiently small (error radius $<10$~arcmin);
\item altitude above horizon is $>20^{\circ}$;
\item the position does not match any known X-ray catalog source;
\item in the case of {\em Swift} alert, the image significance should not be low
(to-date, no real GRB was found to have the low image significance flag raised);
\item none of the five known Soft-Gamma ray Repeaters (SGRs), SGR~1900+14,
SGR~1806-20, SGR~1627-41, SGR~0526-66 and SGR~1801-23, lies in the error circle.
\end{itemize}
As noted above, the average delay between the arrival time of the GCN alert and the start
time of observations ranges between 1 and 3~min.

In the case of an SGR lying in the error circle, the GRB TOCA triggers a
SGR follow-up IR observation with the {\em SupIRCam} according to the
strategy defined in an SGR follow-up approved proposal led by the LT GRB group.
Currently, the automatic SGR follow-up program is still under test, so
the presence of an SGR in the error circle merely invalidates the GRB follow-up
observation.

When all of the above conditions are fulfilled, the GRB TOCA repoints
the telescope, then triggers the acquisition of the first set of
images scheduled in the detection mode (following commissioning of
RINGO (2006), the detection mode exposures will be preceded by
an initial short polarimetric exposure).  The {\em LT-TRAP} is then
invoked to analyze these images and consequently choose the most
suitable observing strategy (see Sec.~\ref{sec:DM} for a detailed
description). The RINGO exposure will be analyzed offline at a later time
and plays no part in real-time decision making; the motivations of
this choice are explained in Section~\ref{sec:PM}.

The GRB TOCA and the {\em LT-TRAP} communicate through the noticeboard,
which is a plain text file continuously updated until the end of observations.
This not only reports in real time a detailed chronological sequence of
all the actions and results acquired through, following technical conventions,
but also represents a useful real-time human-readable interface for
the users responsible for monitoring the on-going observation sequence.

\section{DETECTION MODE (DM)} 
\label{sec:DM} 

The DM consists of a sequence of two sets of three
10-s exposure images taken in $r'$. These are the first images of the
GRB field acquired by the telescope. Once each set has been acquired
and copied to the proxy machine, the GRB TOCA invokes the GRB
astrometric fit script on each image. After object extraction and
comparison with standard catalogs, an OT candidates list is
extracted for each set of three images.  These OT candidates are then
processed by the GRB variability script, which aims to identify
sources common to each image set and evaluate possible variability.
Finally the DM reports the best OT candidate (if any) out of the
two merged lists.

\subsection{The GRB Pipeline Astrometric Fit Script}
\label{sec:GRB_pipe_astrometric}
This shell is responsible for fitting the astrometry and identifying possible
OT candidates. Below we report the detailed
sequence of operations performed on each single image by this script:

\begin{enumerate}
\item Source extraction using a customized version of {\em SExtractor~2.3.2} \citep{Bertin96} is performed;
if the number of sources extracted is lower than 3, it does not attempt to fit the astrometry.
In contrast, if more than 20 sources are extracted, only the 20 brightest ones are used in the astrometric fit procedure.
\item A $9'\times9'$ portion of the USNO--B1.0 astrometric catalog \citep{Monet03}, centered on the
centroid position of the GRB, is retrieved from a complete copy stored locally on the LT proxy
machine. In case of problems accessing the local catalog, the script attempts to retrieve it through the
on-line interface of the USNO--B1.0\footnote{http://www.nofs.navy.mil/data/fchpix/}.
\item The retrieved catalog image is then cropped to match exactly the field of view
of the LT.
\item Optical extinction for the field is determined from the dust maps by
\citet{Schlegel98}.
\item An astrometric fit is performed using an adapted version of the WCS \citep{Greisen02} tool
{\em imwcs~3.5.3} \citep{Mink02}. The fit is considered to have failed if either the
number of matched sources is lower than 3 or if the fraction of matched sources
is lower than 50\%.
\item If the astrometric fit is successful, the pixel-to-sky coordinates transformation
is performed through the WCS tool {\em xy2sky~3.5.7} \citep{Mink02}, based on the result of the
astrometric fit. A typical value for the fit residual is $\sim$0.3~arcsec.
\item In case of astrometric fit failure, all the image sources are considered
in principle as potential OT candidates and they are reported in the OT candidates files,
although with a lower confidence level (see below).
The pixel-to-sky coordinate transformation is performed assuming the
nominal pointing (typical accuracy currently of $\sim$10--20~arcsec, with a target on
completion of commissioning of the telescope of 2 arcsec). In this case, the script
ends at this point.
\item Selection of the OT candidates: if the astrometric fit is successful, all the
image sources are cross-checked with the catalog stars and classified according to
a set of criteria. A 17-bit variable is assigned to each source and describes its
properties. Each image source not associated with any USNO--B1.0 star is assigned 50 points.
Each bit of the 17-bit variable is assigned a given number of points to be added to the
initial 50 points credit (see Table~\ref{tab:GRBpipe}).
\item Selected classes of non-associated image sources are a priori excluded from the
candidate search. Among them, the objects with clearly non-stellar PSFs are probably
the most important (e.g. cosmic ray events, reflections, failed deblending). A mean
stellar PSF is defined on each image based on real-time associated
objects median statistics.
\item Each selected OT candidate is assigned a confidence level directly related to the
total number of points accrued.
\item Each selected OT candidate is cross-checked with the on-line GSC2.3 and 2MASS catalogs
and its confidence level is adjusted accordingly (see Table~\ref{tab:GRBpipe}).
\end{enumerate}

When the astrometric fit is successful, magnitudes are estimated from the USNO--B1.0
stars that have been matched. We use as zero point the median value of the offset needed
to have a match between the instrumental magnitude and the $R$ magnitude reported
in the USNO--B1.0 catalog. The error associated with the zero-point is then evaluated
in terms of the clipped ($2\sigma$) root mean square deviation with respect to the USNO--B1.0.

\begin{deluxetable*}{rrllr}
\tabletypesize{\scriptsize}
\tablecaption{{\em LT-TRAP}: 17-bit variable assigned to each OT candidate\label{tab:GRBpipe}}
\tablewidth{0pt}
\tablehead{
\colhead{Bit $n$} & \colhead{$2^{n-1}$} & \colhead{Description} & \colhead{Script} & \colhead{Points}\\
}
\startdata
 1 &     1 & the extracted magnitude is biased by bright neighbours & Astrometric &   0\\
 2 &     2 & the object has been deblended                          & Astrometric & -10\\
 3 &     4 & at least one pixel is saturated ($>$ 60000 counts)     & Astrometric &   0\\
 4 &     8 & object truncated (close to the image border)           & Astrometric & -50\\
 5 &    16 & aperture data incomplete or corrupted                  & Astrometric & -10\\
 6 &    32 & isophotal data incomplete or corrupted                 & Astrometric & -10\\
 7 &    64 & memory overflow occurred during deblending             & Astrometric & -10\\
 8 &   128 & memory overflow during extraction                      & Astrometric & -20\\
 9 &   256 & wasn't possible to check GSC and/or 2MASS              & Astrometric & -20\\
10 &   512 & a GSC2.3 object lies close ($<$ 2 arcsec)              & Astrometric & -50\\
11 &  1024 & a 2MASS object lies close ($<$ 2 arcsec)               & Astrometric & -30\\
12 &  2048 & a bright USNO star (R$<$13) is close ($<$50px)         & Astrometric & -20\\
13 &  4096 & ASTROMETRIC FIT FAILED  Assuming nominal pointing      & Astrometric & -50\\
14 &  8192 & Object detected in at least 2 OT lists                 & Variability & +30\\
15 & 16384 & Object NOT detected in at least 1 OT list              & Variability & -10\\
16 & 32768 & Variability detected over 2--3 images                  & Variability & +50\\
17 & 65536 & Astrometric fit warning                                & Variability &   0\\
\enddata
\end{deluxetable*}
\subsection{The GRB Pipeline Variability Script}
\label{sec:GRB_pipe_var}
After the {\em LT-TRAP} script described above has run on the three images, the three
resulting OT candidates' files, each reporting a list of potential OTs found in each
single image, are cross-checked through the so-called ``variability script''.
This shell script is invoked by the GRB TOCA and aims to recognize objects common
to each of the OT candidates files, as well as to evaluate possible significant
magnitude variations between the three images. This variation is evaluated by performing
a $\chi^2$ test on the zero hypothesis that there is no significant variation.
If the three values of magnitude are found to be consistent with a constant value
with a significance lower than 1\%, the variability bit is raised (Table~\ref{tab:GRBpipe}).
Each OT candidate confidence level is finally adjusted on the basis of the variability
results  (bit number 16, see Table~\ref{tab:GRBpipe}). 

Table~\ref{tab:GRBpipe} reports the meaning of each bit of the 17-bit variable assigned
to each OT candidate and the points assigned to each flag.
The confidence level is evaluated by dividing the total points by 100: when this ratio is $<0$,
the confidence level is set to zero; when the ratio is $>1$, the level is assigned 1.0
(highest confidence level possible).
The final threshold on the confidence level for an OT candidate to trigger the
Multi-Color Imaging Mode (Sec.~\ref{sec:MCIM}) is currently set at 0.7.
In case of more than one OT candidate
with confidence level above the threshold, the GRB TOCA accepts that with the
highest value. If the ambiguity is still unresolved, it assumes the first on the
list. However, all the information regarding the other candidates is reported in
the noticeboard file as well as in the product files.

The meaning of each single bit described in Table~\ref{tab:GRBpipe} is self-explanatory,
apart from the last bit, called ``Astrometric fit warning'', which requires an explanation.
This bit is raised whenever at least one of the three images has a failure in the
astrometric fit and at least another one has a successful astrometric fit.
In this case, the identification of the same object in different images is performed
on the basis of pixel coordinates, while the sky coordinates are taken from the
image with the correct astrometry. However, so far this case has turned out to
be very rare: either all or none of them are successfully astrometrically fitted.
For this reason, the current points assigned to this bit are zero.

To test the sensitivity of the script, we simulated on real images
an OT fading according to a power law,
$F\propto(t-t_0)^{-\alpha}$ (where $t_0$ is the GRB onset
time). The script detected significant variability for $\alpha\sim$1--2, for
objects at least as bright as $r'\simeq$16--17, provided that the DM
is triggered at $(t-t_0)<200$~s.
Also, in the case of GRB~050502a, where the DM was triggered at $t-t_0=187$~s,
the {\em LT-TRAP} detected significant variability:
the magnitudes evaluated by the {\em LT-TRAP} were for the three DM
images: $r'_1=15.57\pm0.03$,$r'_2=15.66\pm0.03$ and $r'_3=15.83\pm0.03$.
The corresponding $\chi^2$ test gave the following result:
$\chi^2/{\rm dof}=36.2/2$, with a significance of $1.4\times10^{-8}$.

It must be pointed out that the systematic error affecting the USNO--B1.0
magnitudes is around 0.3~mag; furthermore, it is not possible to
correct for color terms at this stage. However, this is not relevant, as
the purpose of the script is to establish relative variations, regardless of
the absolute values. To account for possible variations in the zero point, the
script adopts the following approach: it calculates the root
mean square deviation of the three zero points and adds it in quadrature
to the statistical error of each of the three magnitudes.

\begin{figure*}
\plotone{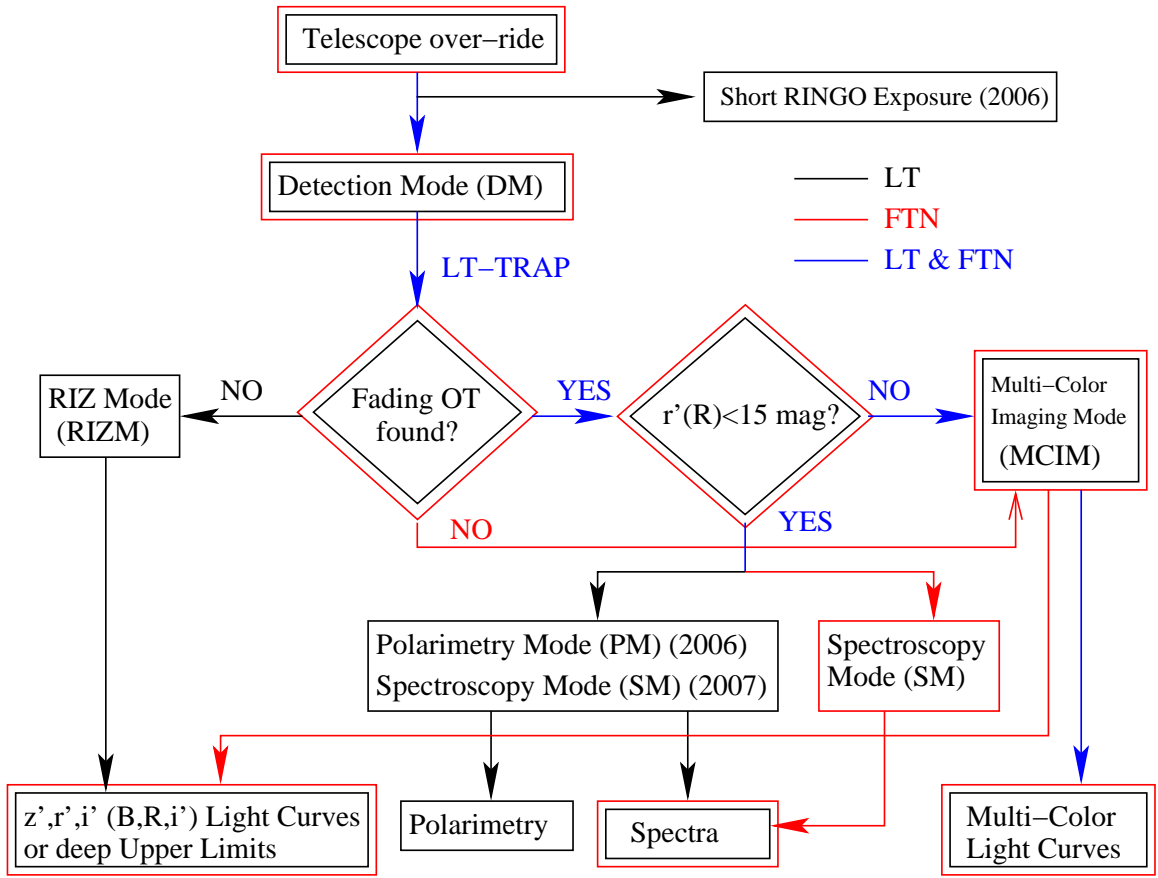}
\caption{Diagram of the {\em LT-TRAP} observing strategy for both LT and FTN facilities.
\label{fig:strategy}}
\end{figure*}

\section{OBSERVATION MODES}

After the DM has completed its imaging and analysis sequence, the
appropriate observation mode is triggered automatically on the basis
of the results of the analysis performed by the {\em LT-TRAP} on the
DM images (see Fig.~\ref{fig:strategy} and Table~\ref{tab:OM}).

\subsection{Multi-Color Imaging Mode (MCIM)}
\label{sec:MCIM}
This mode is triggered by the GRB TOCA when the DM finds  an OT candidate
with the required confidence level.
It consists of a multi-color imaging sequence cycling through three Sloan $g'$,
$i'$ and $r'$ filters until the end of the first
hour.  The ordered sequence, $g'$, $i'$ and $r'$, is repeated with different
exposure times: the first two times with 30-s exposure, then it continues with
60-s exposure cycles up to 18 times.
Before May 16, 2005, this mode operated in the same way but used a slightly
different set of filters, $BVr'i'$, as used for GRB~050502a \citep{Guidorzi05a}.
The motivation for changing and optimizing the filter choice (post May 16) was threefold:
1) the greater sensitivity at $g'$ band, 2) the small separation between $B$ and
$V$ filter central wavelengths ($444.8$~\AA\  and $550.5$~\AA, respectively) and
3) the recognized need for denser sampling of the light curves at early times.

\subsection{Polarimetry Mode (PM)}
\label{sec:PM}

A key unsolved problem in the astrophysics of GRBs is the geometry of
the ejecta and the role of magnetic fields. Direct measurements of
polarization at {\em early times} provide stringent constraints on
current jet and internal shock models. As described in Section
\ref{sec:LT}, we will have unique access to a simple optical
polarimeter, RINGO, on the LT from the end of 2006.  Polarimetry Mode
(PM) will be triggered following the identification by {\em LT-TRAP} of a
candidate OT with $r'<15$~mag and high significance, and will
consist of an automatic series of polarimetric exposures taken with
RINGO, with increasing exposure times:  30s, 60s, 120s, 240s,
etc. These exposures will not be reduced and analyzed in real-time,
but post-processing will extract the $r'$ band light curve and any
polarization evolution of the OT. Although less sensitive than RATCam,
RINGO will detect polarization to a few \% in $r'\lesssim$ 15 mag OTs
in a 30-s exposure. In combination with the initial polarimetric
exposure that will be obtained immediately before the DM begins, this
mode aims to measure the polarization properties from the first
minutes after the burst.

\subsection{Spectroscopy Mode (SM)}
\label{sec:SM}

Following commissioning of FRODOSpec (2007), the identification of a
candidate OT with $r'<15$~mag and high significance will result in
the triggering of an automatic spectroscopic observation. The goal of
the initial spectroscopic observations will be to obtain an early
redshift estimate, whilst subsequent spectra taken during the first
hour will provide valuable information on the time evolution of the
GRB environment and light curve. Short exposure $r'$ will be
interspersed with spectroscopic observations to ensure the OT has not
faded below the sensitivity limit of the spectrograph. In the case of
a rapidly fading OT, the observing mode will revert to MCIM after the
initial spectrum has been obtained.
\begin{deluxetable*}{lll}
\tabletypesize{\scriptsize}
\tablecaption{{\em LT-TRAP Strategy}: observing modes for different facilities\label{tab:OM}}
\tablewidth{0pt}
\tablehead{
\colhead{Observing} & \colhead{Liverpool Telescope} & \colhead{Faulkes Telescope North}\\
\colhead{Mode} & \colhead{} & \colhead{}\\
}
\startdata
DM    & 2 x (3x10 s in $r'$) & (3x10 s in $R$) + (1x10 s in $B, V, i'$)\\
RIZM\tablenotemark{a}   & ($n$ x 2 min) in ($r'$,  $i'$, $z'$)\tablenotemark{a},  $n=1, 2, 3,$\ldots & --\\
MCIM  & 2 x (30 s in $g', r', i'$) + $n$ x (1 min in $g', r', i'$) &
cycles of $B, R, i'$: 1x30s, 1x1min, $n$x (2 min, 3 min)\\
SM    & For OTs with $r'<15$~mag (expected in 2007) & $R<15$~mag\\
PM    & 30s, 60s, 120s... for OTs with $r'<15$~mag (expected in 2006) & --\\
\enddata
\tablenotetext{a}{Original RIM amended to RIZM by inserting additional $z'$ exposures at the end of each $r'i'$ cycle (from  November 2005).}
\end{deluxetable*}

\subsection{RIZ Mode (RIZM)} 
\label{sec:RIZ} 

When the DM does not identify an OT candidate, the GRB TOCA invokes
the ``RIZ Mode''.  This consists of a sequence of 2-min exposure $r'$,
$i'$ and $z'$ images: first, one $r'$, one $i'$ and one $z'$,
then two $r'$ followed by two $i'$ and by two $z'$, then three
$r'$ followed by three $i'$ and three $z'$ and so on. Prior
to November 2005, this mode consisted of $r'i'$ cycles only; we
have subsequently adapted this mode to include additional $z'$ exposures
that are inserted at the end of each $r'i'$ cycle to enable
identification of possible higher redshift bursts (z $\lesssim$ 6)
that may be faint in $r'$ but brighter in $z'$.

This mode is designed to be sensitive to possible faint optical
afterglows (whether intrinsically faint or at higher redshift),
allowing $r', i', z'$ band light curves to be constructed when the
OT is detectable in longer exposures but too faint to have been
identified automatically by the detection mode, or to provide deep
upper limits at early time when no OT is present. Under average
conditions, i.e. seeing of $\sim1$~arcsec, 2-min exposure limiting
magnitudes are around $r'\sim$21--22, $i'\sim$20--21 and
$z'\sim$20.

\section{{\em LT-TRAP} DEPLOYED ON OTHER FACILITIES} The LT is the prime
member of the global 2-m robotic telescopes network called ``{\em
RoboNet-1.0}'' \footnote{See http://www.astro.livjm.ac.uk/RoboNet}.  This
network includes two other facilities: the Faulkes Telescope North
(FTN) located in Maui (Hawaii) and the Faulkes Telescope South (FTS)
in Siding Spring (Australia), both mainly supported by the Dill
Faulkes Educational Trust and intended for use by UK schools and
various scientific observing programs.  All the three telescopes are
operated remotely through the Operations Management Centre in Liverpool
JMU and, when fully commissioned, will all operate in fully robotic
mode.  Under the {\em RoboNet-1.0} project, comprising a consortium
of ten UK universities funded by the UK PPARC,
a fraction of time on the two FTs is reserved for post first hour
GRB follow-up by consortium members (see \citet{Gomboc04,Gomboc05a} for details).

Following the successful operation of {\em LT-TRAP} on the LT, it was
deployed on the Faulkes Telescopes, with a slightly-modified observing
mode choice following the OT identification stage (see Fig.~\ref{fig:strategy}).
FTS is currently undergoing commissioning so is not yet available
robotically; the FTN is fully robotic and has performed a number of
follow-up observations minutes after a GRB; for this reason, hereafter
we will limit our discussion to FTN.  The rapid ($<1$~hr) robotic
follow-up observations of GRBs on FTN is a collaborative project
between Liverpool John Moores University and the University of
Leicester.

The optical filters available on the FT are different from the LT: Bessell
$BVR$, Sloan $i'$, neutral $U$, $O$III and $H\alpha$.
The DM consists of only one set of three 10-s
images in $R$.  Regardless of the results of the DM, three more 10-s
images are acquired in the following filters, respectively: $B$, $V$,
$i'$.  In the case of identification of an OT candidate with $R<15$,
the GRB TOCA will triggers a spectroscopic observation.  As in the case
of LT, the threshold on the confidence level for an OT candidate to be
considered as a good candidate is currently set to 0.7.  The logic of
the selection of the OT candidate is exactly the same as for the LT
(Sec.~\ref{sec:GRB_pipe_var}). Presently, regardless of the possible
identification of an OT candidate with $R>15$ and confidence level
$>0.7$ from the DM, the GRB TOCA triggers a sequence of multi-color
observations according to the following scheme: cycles of $BRi'$ with
increasing exposure times: 30~s (once), 60~s (once), then it goes on
continuously switching between 120~s and 180~s exposures in order to
limit the effects of tracking problems.

\section{Estimation of the {\em LT-TRAP} reliability}
\label{sec:reliability}
\begin{figure}
\plotone{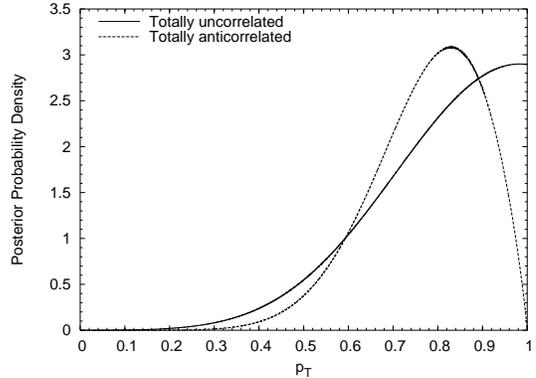}
\caption{Probability density function of $p_{\rm T}$ assuming a truncated
normal prior, $N(0.8, 0.3)$, in two cases: $p_{\rm T}$ and $p_{\rm F}$
uncorrelated (solid line) and totally correlated (dashed line) (see text).
\label{fig:pp}}
\end{figure}

Although the number of GRBs so far observed is still small,
we have attempted to estimate the reliability of the OT candidates
automatically identified by the {\em LT-TRAP} as follows.
For each of the GRBs followed up robotically within minutes,
we counted the OT candidates identified as a function of their
confidence level. When the confidence level is above the threshold
(currently set to 0.7), the OT is taken to be a true OT candidate
(TOC), otherwise it is a fake OT candidate (FOC).
When a true afterglow is present, we call it a true OT (TO);
similarly, if any non-GRB source that could mimic a GRB afterglow is
present, we call it a fake OT (FO). Let $p_{\rm T}$ be the 
probability that a TO is automatically identified as a TOC.
The closer $p_{\rm T}$ to 1, the better the {\em LT-TRAP}
capability of identifying genuine GRB afterglows.
Likewise, let $p_{\rm F}$ be the probability that a FO
is automatically identified as a FOC.
The closer $p_{\rm F}$ to 1, the better the {\em LT-TRAP}
capability of rejecting fake GRB afterglows.
In practice we do know that $p_{\rm T}$ and $p_{\rm F}$ are somehow
correlated: a low threshold on the confidence level would
increase $p_{\rm T}$ but also decrease $p_{\rm F}$.
The opposite is true for a high value of the same threshold.
So far, the current value (0.7) has turned out to be a good
trade-off. However, as a first approximation we treat them as
unrelated and eventually we discuss the opposite case of
total correlation.
Let $n_{\rm T}$ and $m_{\rm T}$ the total number of TO observed
and the total number of TO identified as TOC, respectively.
Likewise, let $n_{\rm F}$ and $m_{\rm F}$ the total number of FO
observed and the total number of FO identified as FOC, respectively.
The probability $P_{\rm T}(n_{\rm T}, m_{\rm T}|p_{\rm T})$
($P_{\rm F}(n_{\rm F}, m_{\rm F}|p_{\rm F})$) to identify
$m_{\rm T}$ ($m_{\rm F}$) TOC (FOC) out of $n_{\rm T}$ TO
($n_{\rm F}$ FO), given $p_{\rm T}$ ($p_{\rm F}$),
is given by the binomial distribution (under the sensible
assumption that different GRBs are independent from each other):
\begin{equation}
P_i(n_i, m_i|p_i) \ = \ {n_i \choose m_i}
\ p_i^{m_i} \ (1-p_i)^{n_i - m_i}\quad ,(i={\rm T}, {\rm F})
\label{eq:binomial}
\end{equation}
We do not know either $p_{\rm T}$ or $p_{\rm F}$, but so far
we counted the following: $n_{\rm T}=2$, $m_{\rm T}=2$;
$n_{\rm F}=15$, $m_{\rm F}=13$ (in fact {\em LT-TRAP} so far
correctly identified both $r'<19$ OTs and failed to reject
2 out of 15 fake OTs).
From the Bayes theorem, we derive the probability
$P_i(p_i|n_i, m_i)$ to have $p_i$, given $n_i$ and $m_i$:
\begin{equation}
P_i(p_i|n_i, m_i) \ = \ \frac{P_i(n_i, m_i|p_i)\ P(p_i)}{P(n_i, m_i)}
\quad ,(i={\rm T}, {\rm F})
\label{eq:bayes}
\end{equation}
We then apply the maximum likelihood method and find the value
for $p_i$ which maximizes $P_i(p_i|n_i, m_i)$. To this end,
the denominator of eq.~\ref{eq:bayes} can be ignored.
Concerning the knowledge of the prior $P(p_i)$, in both cases
as a first guess we may assume the most simple case of a
uniform distribution. Then we refine it thanks to the
information derived from our tests, according to which
a sensible approximation is given by a truncated normal
distribution centered on 0.8 and sigma of 0.3.
However, it must be pointed out that the prior strongly depends
on the kind of afterglow population one considers.
For this reason, we consider these two opposite cases
of uniform and truncated normal distributions.
When a uniform prior is assumed, the best value is straightforwardly
given by $p_i=m_i/n_i$: $p_{\rm T}=2/2=1$ and
$p_{\rm F}=13/15\simeq 0.87$.
When we assume a more refined prior, i.e. $P(p_i)\sim N(0.8, 0.3)$,
the best values are $p_{\rm T}\simeq 0.98$ (Fig.~\ref{fig:pp}, solid line)
and $p_{\rm F}\simeq 0.86$.
Eventually, we can provide confidence intervals on both $p_{\rm T}$ and
$p_{\rm F}$ at given levels from the distributions obtained with
the truncated normal priors:
$p_{\rm T}=0.98_{-0.16}^{+0.02}$ and
$p_{\rm F}=0.86_{-0.06}^{+0.05}$ (50\% CL);
$p_{\rm T}=0.98_{-0.42}^{+0.02}$ and
$p_{\rm F}=0.86_{-0.17}^{+0.10}$ (90\% CL).

If we take into account the correlation between $p_{\rm T}$ and
$p_{\rm F}$, the total probability of eq.~\ref{eq:bayes} is replaced
by the more general one:
\begin{eqnarray}
P_{\rm T,F}(p_{\rm T},p_{\rm F}|n_{\rm T}, m_{\rm T}, n_{\rm F}, m_{\rm F})
\ =~~~~~~~~~~~~~~~~~~~~~~~\nonumber\\
~~~~~\ \frac{P_{\rm T}(n_{\rm T}, m_{\rm T}|p_{\rm T})\
P_{\rm F}(n_{\rm F}, m_{\rm F}|p_{\rm F})\ P(p_{\rm T}, p_{\rm F})}{
P_{\rm T}(n_{\rm T}, m_{\rm T}, n_{\rm F}, m_{\rm F})}
\label{eq:bayes2}
\end{eqnarray}
At this stage, it is difficult to model the anticorrelation between
$p_{\rm T}$ and $p_{\rm F}$ in the bivariate prior $P(p_{\rm T},p_{\rm F})$,
particularly for the small sample of GRBs so far collected.
We then studied the opposite case of a total anticorrelation between
the two variables, in which $p_{\rm F}$ is completely determined by
the value of $p_{\rm T}$, by solving eq.~\ref{eq:anticorr}:
\begin{equation}
\int_{p_{\rm F}}^{1} P_{\rm F}(p'|n_{\rm F}, m_{\rm F})\,dp'\ =\ 
\int_{0}^{p_{\rm T}} P_{\rm T}(p'|n_{\rm T}, m_{\rm T})\,dp'
\label{eq:anticorr}
\end{equation}
where $P_i(p'|n_i, m_i)$ is the same expression derived in
eq.~\ref{eq:bayes}, assuming truncated normal priors.
The meaning of eq.~\ref{eq:anticorr} is that, for each value
$p_{\rm T}'$, $p_{\rm F}'$ is determined
so that the probability that $p_{\rm F}>p_{\rm F}'$, is equal
to the probability that $p_{\rm T}<p_{\rm T}'$; it is straightforward
to verify that the greater $p_{\rm F}'$, the smaller $p_{\rm T}'$
and vice versa. This is just one sensible assumption of total
anticorrelation among other possible alternative assumptions.
Figure~\ref{fig:pp} shows the probability density for $p_{\rm T}$
derived in this case of total anticorrelation (dashed line).
The best values for $p_{\rm T}$ and $p_{\rm F}$ and their confidence
intervals derived in this case (truncated normal priors) are the
following:
$p_{\rm T}=0.83_{-0.09}^{+0.08}$ and
$p_{\rm F}=0.83_{-0.05}^{+0.04}$ (50\% CL);
$p_{\rm T}=0.83_{-0.24}^{+0.14}$ and
$p_{\rm F}=0.83_{-0.24}^{+0.09}$ (90\% CL).

Remarkably, in spite of the opposite cases of variables
$p_{\rm T}$ and $p_{\rm F}$ either totally unrelated or totally
anticorrelated, the results seem to give consistent confidence
intervals, as shown by Fig.~\ref{fig:pp}.
The wide confidence interval of $p_{\rm T}$ is understandably due to
the low number of TO observed so far, rather than the limits of the
{\em LT-TRAP}.
We expect to give more precise estimates of both probabilities
as soon as the sample of GRB afterglows detected is more
numerous.

\section{SOME RESULTS AND FUTURE PLANS}
\label{sec:results}
To-date (October 2005), over a period of two months, LT followed up in a
fully automated fashion using the real-time pipeline six GRBs and FTN followed
up five more. Table~\ref{tab:results} reports for each of these
eleven follow-up observations the start time from the GRB trigger,
the filters used, the number of frames acquired and the GCN circulars
issued by the LT/FTN collaboration reporting on the corresponding LT/FTN
observations.
The most successful case so far is represented by the automatic detection
of the afterglow of GRB~050502a \citep{Gomboc05b} followed by the
acquisition of the first early ($<1$hr) multi-color light curve
\citep{Guidorzi05a}. This latter case is also notable for longer term follow-up using
the {\em RoboNet-1.0} network. The other case of prompt detection is GRB~050713a
\citep{Monfardini05a}, although in this case the afterglow was not identified
automatically because of the poor quality of the images affected by
the presence of a $V\sim6$ star in the field, but recognized afterward
by visual inspection.
\begin{deluxetable*}{llllrlrc}
\tabletypesize{\scriptsize}
\tablecaption{Results of the robotic follow-up observations with Liverpool and Faulkes North Telescopes (October 2005)\label{tab:results}}
\tablewidth{0pt}
\tablehead{
\colhead{Telescope} &\colhead{GRB} & \colhead{Spacecraft} &  \colhead{Filters} & \colhead{Start}\tablenotemark{a} & \colhead{Result} & \colhead{Frames} & \colhead{LT/FTN GCN}\\
              &                      &                     &                   & \colhead{(min)} &                &   & \colhead{Circulars}\\
}
\startdata
FTN & 050412  & {\em Swift}    & $BVRi'$   &  2.5  & $R>18.7$     &  23  & none\\
LT  & 050502a & {\em INTEGRAL} & $BVr'i'$  &  3.1  & $r'\sim15.8$ &  25  & 3325\\
FTN & 050504  & {\em INTEGRAL} & $BVRi'$   &  3.7  & $R>19$       &  38  & 3351\\
LT  & 050520  & {\em INTEGRAL} & $r'i'$    &  4.5  & $r'>16.6$    &  18  & 3437\\
LT  & 050528  & {\em Swift}    & $r'i'$    &  2.5  & $r'>17.2$    &  36  & 3497\\
LT  & 050713a & {\em Swift}    & $r'$      &  2.4  & $r'\sim19.2$ &   3  & 3588\\
FTN & 050713b & {\em Swift}    & $R$	   &  3.3  & $R>18.2$     &   6  & 3592\\
FTN & 050716  & {\em Swift}    & $BVRi'$   &  3.8  & $R>19.8$     &  20  & 3625\\
LT  & 050730  & {\em Swift}    & $r'i'$    &   50  & $r'\sim17.3$ &  36  & 3706\\
LT  & 050904  & {\em Swift}    & $r'i'$    &  3.8  & \tablenotemark{b} & 31 & none\\
FTN & 050925  & {\em Swift}    & $BVRi'$   &  3.3  & $R>19.0$     &  12  & 4035\\
\enddata

\tablenotetext{a}{This corresponds to the time delay with
respect to the GRB trigger time.}
\tablenotetext{b}{The afterglow of GRB050904 \citep{Haislip05} 
was found to lie outside the field of view of the LT (3.9~arcmin away
from the BAT position).}.
\end{deluxetable*}
The power and robustness of the {\em LT-TRAP} is illustrated in the case
of GRB~050730 \citep{Gomboc05c}: this GRB occurred before twilight at
the LT site, so was triggered manually and, although the OT was
not recognized during the DM because of the high sky background
due to the sky not yet being completely dark, it was automatically identified 
in the subsequent RIM as an unknown fading source of $r'=17.3$ at
$t\sim50$~min after the GRB.
In the other cases, the combination of sensitivity and rapid response
made it possible to derive deep upper limits, as in the cases of
GRB~050504 with $R>19$ \citep{Monfardini05b} and GRB~050716 with
$R>19.8$ \citep{Guidorzi05b} 3--4~min after the GRB trigger time.

In the future, we plan to refine the strategy when an OT candidate
with $R>15$ and confidence level $>0.7$ is found from the DM; in
particular, exposure times will be tuned dynamically on the basis of
the estimated magnitude of the best OT candidate. Automatic
 polarimetry, near-infrared and spectroscopy follow-up modes will be
implemented on the LT following completion of instrument commissioning.

\acknowledgments
CG and AG acknowledge their Marie Curie Fellowships from the European Commission.
CGM acknowledges financial support from the Royal Society.
AM acknowledges financial support from the UK PPARC.
MFB is supported by a PPARC Senior Fellowship.
The Liverpool Telescope is operated on the island of La Palma by
Liverpool John Moores University at the Observatorio del Roque de los
Muchachos of the Instituto de Astrofisica de Canarias.
The Faulkes Telescope North is operated with support from the Dill
Faulkes Educational Trust.

\end{document}